# Alternative interpretation of the recent experimental results of angle-resolved photoemission spectroscopy on GaMnAs [Sci. Rep. 6, 27266 (2016)]

Masaki Kobayashi,[1,2,*] Shinobu Ohya,[3,4] Iriya Muneta,[5] Yukiharu Takeda,[6] Yoshihisa Harada,[7] Juraj Krempaský,[1] Thorsten Schmitt,[1] Masaharu Oshima,[8] Vladimir N. Strocov,[1] Masaaki Tanaka,[3,4] and Atsushi Fujimori[9]

[1]*Swiss Light Source, Paul Scherrer Institut, CH-5232 Villigen PSI, Switzerland*

[2]*Department of Applied Chemistry, School of Engineering, University of Tokyo, 7-3-1 Hongo, Bunkyo-ku, Tokyo 113-8656, Japan*

[3]*Department of Electrical Engineering and Information Systems, The University of Tokyo, 7-3-1 Hongo, Bunkyo-ku, Tokyo 113-8656, Japan*

[4]*Center for Spintronics Research Network, Graduate School of Engineering, The University of Tokyo, 7-3-1 Hongo, Bunkyo-ku, Tokyo 113-8656, Japan*

[5]*Department of Electrical and Electronic Engineering, Tokyo Institute of Technology, 4259 Nagatsuta, Midori-ku, Yokohama, Kanagawa 226-8503, Japan*

[6]*Materials Sciences Research Center, JAEA, Sayo, Hyogo 679-5148, Japan*

[7]*Institute for Solid State Physics, The University of Tokyo, Kashiwanoha, Kashiwa, Chiba 277-8561, Japan*

[8]*Synchrotron Radiation Research Organization, The University of Tokyo, 7-3-1 Hongo, Bunkyo-ku, Tokyo 113-8656, JAPAN*

[9]*Department of Physics, The University of Tokyo, Bunkyo-ku, Tokyo 113-0033, Japan*

**Abstract**

Clarification of the position of the Fermi level ($E_F$) is important in understanding the origin of ferromagnetism in the prototypical ferromagnetic semiconductor $Ga_{1-x}Mn_xAs$ (GaMnAs). In a recent publication, Souma *et al*. [Sci. Rep. **6**, 27266 (2016)], have investigated the band structure and the $E_F$ position of GaMnAs using angle-resolved photoemission spectroscopy (ARPES), and concluded that $E_F$ is located in the valence band (VB). However, this conclusion contradicts a number of recent experimental results for GaMnAs, which showed that $E_F$ is located above the VB maximum in the impurity band (IB). Here, we show an alternative interpretation of their ARPES experiments, which is consistent with those recent experiments and supports the picture that $E_F$ is located above the VB maximum in the IB.

There has been a long-standing dispute over the Fermi level ($E_F$) position in the prototypical ferromagnetic semiconductor $Ga_{1-x}Mn_xAs$ (GaMnAs). Determination of the $E_F$ position is particularly important to clarify the origin of the ferromagnetism of GaMnAs. In Ref. 1, Souma *et al.* have investigated the band structure and the $E_F$ position of GaMnAs using angle-resolved photoemission spectroscopy (ARPES). They have concluded that $E_F$ is located in the valence band (VB). In this case, ferromagnetism is thought to be stabilized by the itinerant VB holes interacting with the localized *d* electrons of the doped Mn atoms. (Hereafter, we refer to this picture as the VB conduction picture.) However, this conclusion contradicts previous experiments including soft X-ray ARPES,[2] resonant tunneling spectroscopy,[3–6] magnetic circular dichroism,[7,8] infrared absorption spectroscopy,[9] and optical pump-probe measurements,[10,11] which showed that $E_F$ is located above the VB maximum in the impurity band (IB). (Hereafter, we refer to this picture as the IB conduction picture.) Here, we present an alternative interpretation of the ARPES measurements reported by Souma *et al.*[1] Our interpretation is consistent with the IB conduction picture and likely more reasonable to understand their results than the VB conduction picture.

Ambiguity of the interpretation of their results originates from the following reasons. As can be seen in Fig.1c,d in Ref. 1, the spectral intensity of the VB is strongly suppressed in the energy range from ~1 eV to $E_F$. Thus, it is difficult to unambiguously identify each band in this region. Also, because the second derivative of momentum distribution curves (MDCs) is known to be inapplicable near the top or bottom of energy bands, the top of bands A and B show a Λ-like shape, which is obviously different from the theoretical dispersion curves obtained by the tight binding calculation shown in Fig. 1e in Ref. 1. Note that, in Ref. 1, band A is assigned to the split-off (SO) band of GaMnAs. The SO band was reported to protrude the heavy-hole (HH) and light-hole (LH) bands in the surface-sensitive ARPES measurements by Kanski *et al.*[12] The reason for the large deformation of the SO band is not so clear at present, but the top of the SO band reaches very close to the $E_F$ as in Fig. 2 of Ref. 1. Because such a deformation has not been detected in the bulk state of GaMnAs,[2] this may be caused by unknown surface states. (It should be noted that in Kanski *et al.*'s data[12] the HH and LH bands do not cross $E_F$ unlike Ref. 1.) These results make it questionable to determine the VB position relative to $E_F$ only from Souma *et al.*'s experimental results. Their conclusion that $E_F$ is located in the VB was *not* directly obtained by the measurement results but was derived *from the fitting curves* of the calculated band structure of GaAs obtained by the tight binding model.

Here, we show an alternative way of fitting of the theoretical band structure to their

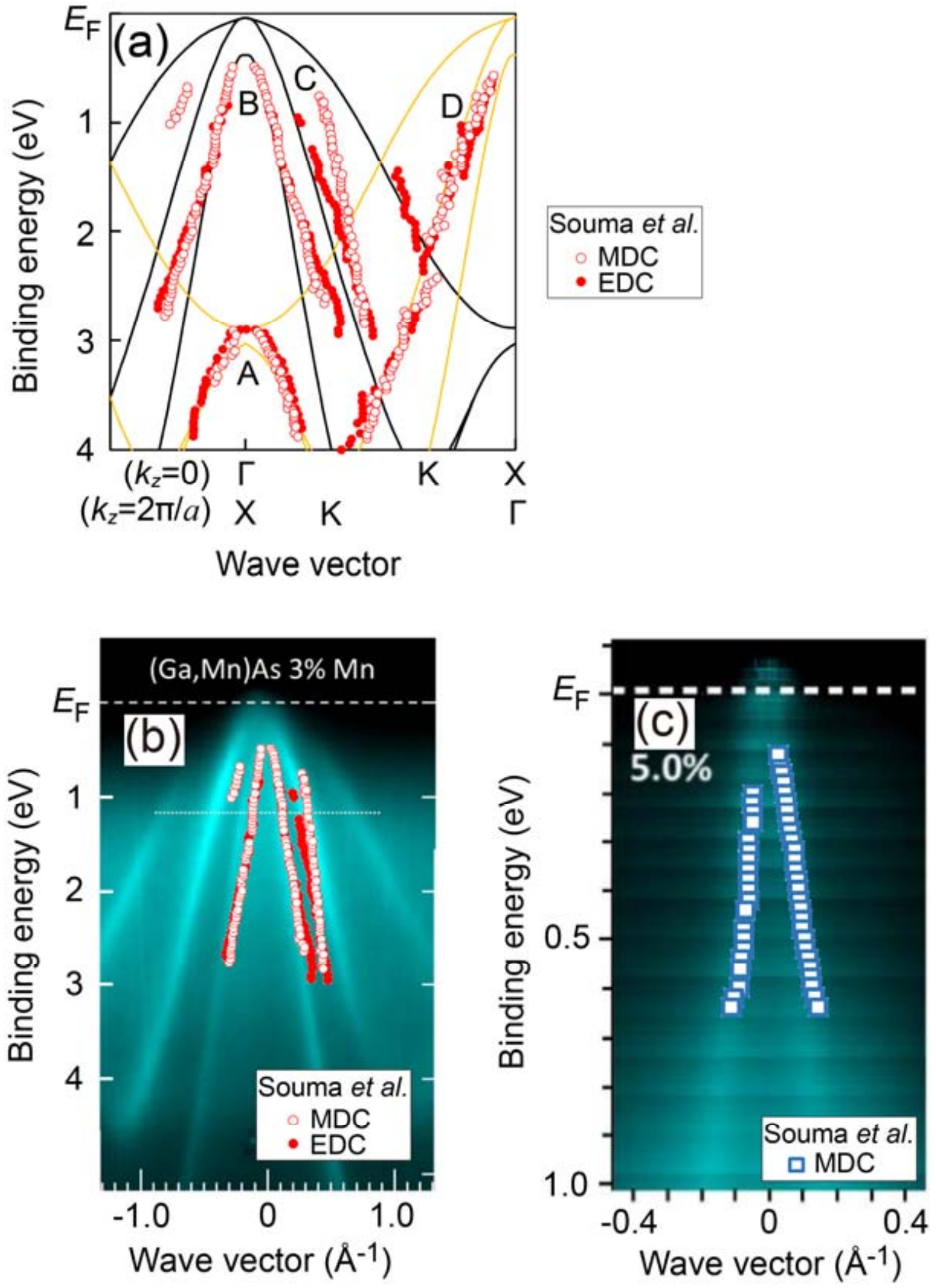

**Fig. 1 (a) Comparison between the ARPES data of Ref. 1 and the calculated band structure of GaAs when $E_F$ is located at 40 meV above the VB maximum (*i.e.* IB conduction picture). The black and orange curves are the calculated band dispersions of GaAs[1] at the wave vector $k_z = 0$ (ΓKX plane) and $k_z = 2\pi/a$ (XKΓ plane). Here, *a* is the lattice constant of GaAs. Bands are labeled with A-D. (b),(c) Color map of the ARPES data[12] obtained with photon energies of 453 eV (b) and 21 eV (c) (courtesy of J. Kanski). The red open circles (or blue open squares) and the filled circles are the second derivative of the intensity of MDCs and the ARPES spectra reproduced from Ref. 1, respectively.**

ARPES data. The red open and filled circles in Fig. 1(a) in this paper show peaks in the second derivative intensity of MDCs and ARPES spectra of $Ga_{0.95}Mn_{0.05}As$ reproduced from Ref. 1. Figure 1 also shows the theoretical band structure of GaAs (black and orange curves) when $E_F$ is located at 40 meV above the VB maximum in the band gap[4] (*i.e.* IB conduction picture). One can see that the theoretical curves well fit to the experimental results, especially to the ARPES spectra (red filled points). This indicates that their ARPES data *do not contradict* the IB conduction picture. Although the experimental data show a small deviation from the theoretical curves at the top of band A, this is probably because the signal is heavily broadened and the spectral shape is strongly deformed as mentioned above. A similar phenomenon of intensity broadening occurs near the top of band B (very faint blue region in Fig. 1d of Ref. 1). In Fig. 2 of Ref. 1, the $E_F$ position was discussed with this largely deformed band B in detail. However, because the intensity shape is largely different from the theoretical dispersion curves near the top of band B, we think that more careful analysis is necessary for the determination of the $E_F$ position relative to the VB. Also, we emphasize again that MDC data sometimes give a wrong band dispersion especially near the top of the VB or the bottom of the conduction band.[13]

Kanski *et al.*[12] reported that the deformed SO band was observed when using a low incident photon energy. In their ARPES data obtained with a high photon energy of 453 eV shown in Fig. 1(b), the HH and LH bands do not cross $E_F$, which is consistent with the IB conduction picture and agrees well with the ARPES data of Ref. 2. This is clearly different from the case of another Mn- and hole-doped semiconductor, where the $E_F$ is located in the VB.[14] The data by Souma *et al.* almost follow Kanski *et al.*'s data [Fig.1(b)]. On the other hand, when using a low incident photon energy of ~20 eV, Kanski *et al.* observed a strongly deformed SO band, which extends upward and approaches $E_F$ as the Mn concentration increases [Fig. 1(c)]. Because Souma *et al.*'s data well follow this deformed SO band reported by Kanski *et al.* [Fig. 1(c)], the strongly deformed bands observed by both groups are likely the same. The important point is that, in Kanski *et al.*'s data, the shapes of the HH and LH bands are not deformed and they remain below $E_F$. As the Mn concentration increases, the relative position between these (HH and LH) bands and the SO band is changed. Thus, it is impossible to estimate the correct $E_F$ position from this strongly deformed SO band using a band structure calculation of *GaAs* without experimental data of the top of the HH and LH bands.[15]

There seems inconsistency between the experimental results shown in Fig. 2d,f,g and their model based on the VB conduction picture shown in Fig. 3b of Ref. 1. Their experimental results showed that the $E_F$ position relative to the VB does not change appreciably when the Curie temperature ($T_C$) decreases from 101 K to 62 K as can be seen in Fig. 2d,f,g of Ref. 1. This result does not agree with the VB conduction picture, in which $E_F$ should strongly move with a decrease in $T_C$ as schematically shown in Fig. 3b of Ref. 1. Their results shown in Fig. 2d,f,g are more likely consistent with the IB conduction picture, in which $E_F$ is pinned in the IB and does not move remarkably with a change in $T_C$.[4] Actually, one can see a feature that may be related to the IB near $E_F$ in their results. In Fig. 2c of Ref. 1, one can see a small peak at the binding energy around 0.08 eV. The position of this peak does not depend on the wave vector $k_x$ (*i.e.* the position is fixed in all curves in Fig. 2c). This is definitely different from the feature of the VB, but may be understood as a feature related to the IB.

We note that GaMnAs is an inherently disordered system where the disorder results in energy broadening of the IB and, although to a lesser extent, of the GaAs host bands.[16] Therefore, in principle, the VB weight could extend to $E_F$ and contribute to the ferromagnetic transport, with the magnitude of this contribution scaling up with the degree of disorder and thus the Mn concentration $x$. However, as can be seen in the previous resonant tunneling studies, quantization of the VB has been clearly observed in $Ga_{1-x}Mn_xAs$ quantum wells with $x$ up to 15%.[3,4,6] This result indicates that disorder is not significant in the VB, and confirms the IB conduction picture.

The most important feature of the VB conduction picture is that the VB, including the HH, LH, and SO bands, is clearly spin split due to the *p-d* exchange interaction. The mean-field *p-d* Zener model, which is a representative VB conduction model, has predicted that the spin splitting is 60 meV for the LH and SO bands when $x$ is 5%.[17] Because the energy resolution of the ARPES measurements was 15–40 meV in Ref. 1, the spin splitting may be observed, if any. In Ref. 1, however, it cannot be seen in any data. In the IB conduction picture, the spin splitting of the VB is estimated to be less than 3 meV.[2,3] Thus, this result is likely consistent with the IB conduction picture.

In summary, the observation of the maximum of the SO band reported in Ref. 1 is unclear, which is not enough to support the VB conduction picture. Meanwhile, one can see features which are well explained by the IB conduction picture in the ARPES measurements reported in Ref. 1. Also, some features of their results seem to contradict the VB conduction picture. Therefore, we think that the ARPES results reported in Ref. 1 likely support the IB conduction picture.

**References**

*Present address: Photon Factory, Institute of Materials Structure Science, High Energy Accelerator Research Organization (KEK), 1-1 Oho, Tsukuba 305-0801, Japan; masakik@post.kek.jp

[1] S. Souma, L. Chen, R. Oszwałdowski, T. Sato, F. Matsukura, T. Dietl, H. Ohno, and T. Takahashi, Sci. Rep. **6**, 27266 (2016).

[2] M. Kobayashi, I. Muneta, Y. Takeda, Y. Harada, A. Fujimori, J. Krempaský, T. Schmitt, S. Ohya, M. Tanaka, M. Oshima, and V. N. Strocov, Phys. Rev. B **89**, 205204 (2014).

[3] S. Ohya, K. Takata, and M. Tanaka, Nature Phys. **7**, 342-347 (2011).

[4] I. Muneta, H. Terada, S. Ohya, and M. Tanaka, Appl. Phys. Lett. **103**, 032411 (2013).

[5] M. Tanaka, S. Ohya, and P. N. Hai, Appl. Phys. Rev. **1**, 011102 (2014).

[6] I. Muneta, S. Ohya, H. Terada and M. Tanaka, Nature Commun. **7**, 12013 (2016).

[7] K. Ando, H. Saito, K. C. Agarwal, M. C. Debnath, and V. Zayets, Phys. Rev. Lett. **100**, 067204 (2008).

[8] H. Terada, S. Ohya, and M. Tanaka, Appl. Phys. Lett. **106**, 222406 (2015).

[9] K. S. Burch, D. B. Shrekenhamer, E. J. Singley, J. Stephens, B. L. Sheu, R. K. Kawakami, P. Schiffer, N. Samarth, D. D. Awschalom, and D. N. Basov, Phys. Rev. Lett. **97**, 087208 (2006).

[10] T. Matsuda and H. Munekata, Phys. Rev. B **93**, 075202 (2016).

[11] T. Ishii, T. Kawazoe, Y. Hashimoto, H. Terada, I. Muneta, M. Ohtsu, M. Tanaka, and S. Ohya, Phys. Rev. B **93**, 241303(R) (2016).

[12] J. Kanski, L. Ilver, K. Karlsson, M. Leandersson, I. Ulfat, and J. Sadowski, arXiv:1608.06821.

[13] S. Aizaki, T. Yoshida, K. Yoshimatsu, M. Takizawa, M. Minohara, S. Ideta, A. Fujimori, K. Gupta, P. Mahadevan, K. Horiba, H. Kumigashira, and M. Oshima, Phys. Rev. Lett. **109**, 056401 (2012).

[14] H. Suzuki, G. Q. Zhao, K. Zhao, B. J. Chen, M. Horio, K. Koshiishi, J. Xu, M. Kobayashi, M. Minohara, E. Sakai, K. Horiba, H. Kumigashira, Bo Gu, S. Maekawa, Y. J. Uemura, C. Q. Jin, and A. Fujimori, Phys. Rev. B **92**, 235120 (2015).

[15] S. Souma, L. Chen, R. Oszwałdowski, T. Sato, F. Matsukura, T. Dietl, H. Ohno, and T. Takahashi, arXiv:1609.01047.

[16] A. X. Gray, J. Minár, S. Ueda, P. R. Stone, Y. Yamashita, J. Fujii, J. Braun, L. Plucinski, C. M. Schneider, G. P`anaccione, H. Ebert, O. D. Dubon, K. Kobayashi, and C. S. Fadley, Nature Mater. **11**, 957 (2012).

[17] T. Dietl, H. Ohno, and F. Matsukura, Phys. Rev. B **63**, 195205 (2001).